\documentclass[twocolumn,floats,10pt,twoside,a4paper]{article}
\usepackage{graphicx}
\usepackage{dcolumn}
\usepackage{amsfonts}
\usepackage{euler}

\def\s2{{\mathcal S}_2} 
\def\wr{{\mathcal W}r}
\def\ts{{\bf t}(s)}
\def\n{{\bf n}} 
\def\t{{\bf t}} 
\def\b{{\bf b}} 

\begin{document}
\title{Writhing Geometry at Finite Temperature:\\
  Random Walks and Geometric phases for Stiff Polymers}
\author{A.C. Maggs. \\ 
  {\it PCT, ESPCI, 10 Rue Vauquelin, Paris, 75005.} 
}
\date{\today}
\maketitle

\begin{abstract}
{\small
  We study the geometry of a semiflexible polymer at finite
  temperatures. The writhe can be calculated from the properties of
  Gaussian random walks on the sphere.  We calculate static and
  dynamic writhe correlation functions. The writhe of a polymer is
  analogous to geometric or Berry phases studied in optics and wave mechanics.
  Our results can be applied to confocal microscopy studies of stiff
  filaments and to simulations of short DNA loops.} 
\end{abstract}


\section{Introduction}

Experiments which micro-manipulate stiff polymers, such as actin
filaments or DNA have lead to a renewed interest in the static and
dynamic properties of semiflexible polymers. This article treats the
writhing properties of stiff polymers, in particular those which have
a length comparable to the persistence length. The static and dynamic
writhing properties discussed here can be easily measured using
confocal microscopy on fluorescently marked actin filaments for which
the persistence length is $15\mu m$, and which are readily available
in a range of length varying between $5\mu m$ to $30 \mu m$.

In this article I concentrate almost exclusively on the writhing
fluctuations of polymers; clearly a stiff polymer has both bending and
twisting degrees of freedom, but it is known \cite{me0,wiggins,nelson}
that the relaxation times of the twist modes are much faster than
those of bending modes, which drive the writhe dynamics. Thus the
longtime effective degrees of freedom of a fluctuating stiff polymer
are those of writhe discussed here.

At finite temperatures the trajectory of a semiflexible filament is
not smooth and direct application of the standard relationships
between, for instance geometric torsion and writhe turns out to be
quite subtle. Indeed the usual definition of torsion requires curves
that are ${\cal C}^3$, whereas the trajectory of a thermalized stiff
polymer is much less smooth than this; in fact both the torsion and
its integral are divergent so that the usual relationship between
writhe and torsion, $2\pi \wr+\int \tau ds\ =0\; mod \ 2 \pi$, is useless.  Here
we explore representations of the trajectory of a polymer based on
random walks on the sphere which have the advantage of being easily
visualized. We are able to use this representation to calculate the
static and dynamic fluctuations of a stiff polymer.  We discuss the
analogy between Fuller's result for the writhe of a polymer and the
geometric phase known from quantum and optical physics.

The main quantitative predictions made in this paper are for the
static and dynamic correlation functions for the writhe, $\wr$ in
several different geometries.  In particular we calculate
\begin{math}
  \langle \wr ^2 \rangle
\end{math} and 
\begin{math}
  \langle ( \wr(t) - \wr(0) )^2 \rangle
\end{math}
for open and closed semiflexible polymers.  With these results we are
able to understand quantitatively a number of results for writhe
autocorrelation functions which have been measured recently in
numerical models of DNA \cite{tamar}.  We also show that short, highly
curved sections of polymer give enhanced contributions to the writhe
of a polymer. We explicitly calculate the probability distribution for
writhing, ${\cal P}_K(\wr)$ for short polymers and show that it is
strongly non-Gaussian.

\section{Writhe of a curve}
\begin{figure}[ht]
  \includegraphics[scale=.22] {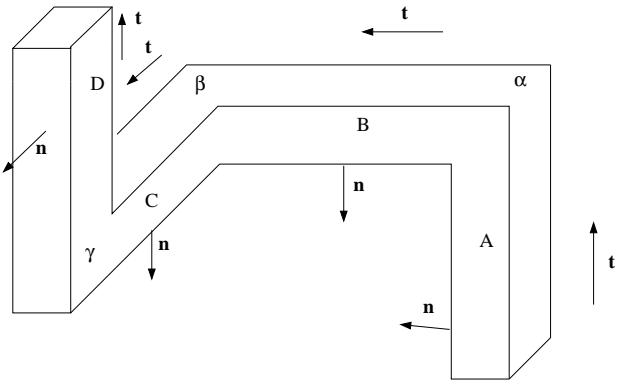}
\caption{
\small {A bar is bent three times, leading to a rotation of $90^\circ$ of
  the vector ${\bf n}$ marking one face of the prism. This rotation is
  due to the writhing of the shape in three dimensions.
 } }\label{writhefig}
\end{figure}

Consider fig. (\ref{writhefig}). It represents a solid bar that has
been bent three times by $90^\circ$.  On one face of the bar I have
added an unit arrow, ${\bf n}$, which is perpendicular to the local
tangent of the bar, ${\bf t}$.  Following the motion of $\n$ we notice
that after the triple bend there has been a rotation of the arrow by
$90^\circ$ about the vertical axis despite the fact that $\t$ has come
back to its original direction.  This rotation is due to the {\it
  writhe}\/ of the path.  Mathematicians measure writhe in units of $2
\pi$. Thus fig. (\ref{writhefig}) corresponds to a writhe, $\wr$ of
$1/4$.

\begin{figure}[ht]
  \includegraphics[scale=.35] {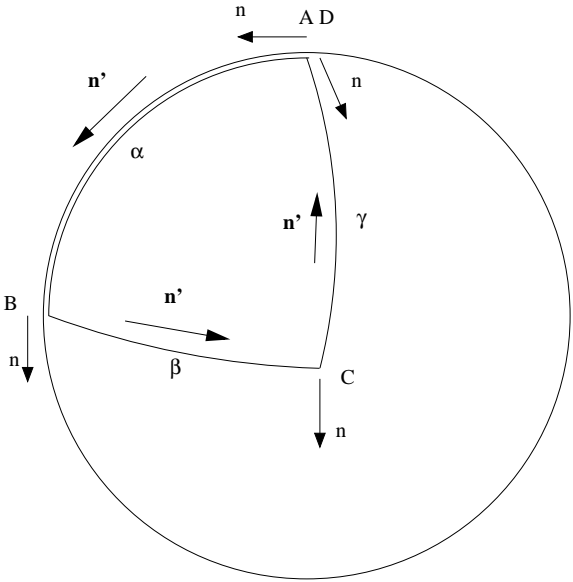}
\caption{{\small Trajectory in the tangent space. The points $A$, $B$, $C$, $D$
    correspond to the straight sections of the prism in fig.
    (\ref{writhefig}). The three bends correspond to the geodesic arcs
    $\alpha$, $\beta$, $\gamma$. The vector ${\bf n}$ is parallel
    transported along the path, leading to an angle of $\pi/2$ between
    the initial and final vectors $\n$.} The vector $\n'$
  (corresponding to the Frenet frame) always points in the direction of
  motion on the sphere.  }
\label{writhefig2}
\end{figure}

The writhing in fig. (\ref{writhefig}) can be best understood by
studying the geometry of the tangent vector ${\bf t}$ together with
the normal vector ${\bf n}$. Since the vector ${\bf t}$ has unit magnitude,
is thus lives on a unit sphere, $\s2$, drawn in figure
({\ref{writhefig2}}). The vector ${\bf n}$ is perpendicular to ${\t}$:
it must always remain tangential to the sphere.  The geometry of the
bent bar in fig. ({\ref{writhefig}}) has been translated onto the
sphere in fig. ({\ref{writhefig2}}). Start at bottom right of
fig. (\ref{writhefig}) and move vertically, following the vector
$\t$.  The whole vertical section, $A$ corresponds to a single point
on the sphere, the north pole. Passing through the first bend $\alpha$
corresponds to the arc, also labeled $\alpha$ on the sphere. The
second straight section of the bar, $B$ points to the left and
corresponds to the point $B$ on the left most edge of the sphere.
Despite the {\it discontinuities}\/ in the direction of bar the vector
${\bf n}$ is always {\it continuous}\/ on $\s2$.  The vector ${\bf n}$
is {\it parallel transported}\/ around a loop on the sphere.

The rotation of the vector ${\bf n}$ is a manifestation of the {\it
  curvature}\/ of the sphere.  Parallel transport of a vector around a
closed path on a surface leads to a angle between the initial and
final vectors.  For a two dimensional surface this angle is given by
\begin{equation}
\Delta_\Omega= \int {\cal K}\ dS \quad modulo \ 2\pi \ ,
\label{gauss}
\end{equation}
where ${\cal K}$ is the Gaussian curvature of the surface and the
integral is over the region enclosed by the path. For a sphere the
Gaussian curvature is equal to unity so that the angle of rotation
between two ends of a bent prism is calculated from the area enclosed
by the trajectory on the sphere. This is just the contents of Fuller's
result \cite{fuller}. In our simple example the solid angle enclosed
by the the curve $(\alpha,\beta,\gamma)$ is $\pi/2$ leading to the
rotation of $\n$ by $90^\circ$.

Any real prism, made of an elastic material has an additional degree
of freedom associated with it: At any point it can be twisted, at
angular velocity $\omega_{\t}(s)$ about the tangent, compared with the
parallel transported frame shown in figures
(\ref{writhefig},\ref{writhefig2}).  Clearly the total rotation in the
laboratory frame is given as the sum of the rotation due to the
writhing, plus the rotation $\int \omega_{\t} \ ds$ coming from the
relative motion of the parallel transported and twisting frame. This
is the content of White's theorem \cite{white} for a curve. ${\cal L}
n={\cal W }r+{\cal T}w$, where $2\pi {\cal L}n$ is the total rotation,
$2\pi {\cal T}w$ the rotation due to twisting and $2\pi \wr$ the
rotation due to writhing.

\section{Geometry of three dimensional curves}

In this section I shall review some of the elementary properties of a
line in three dimensions, pointing out the relationships between
different representations of the geometry. In particular I will
discuss the classic choice of the Frenet frame and compare it
with the parallel transported frame of fig. (1,2) to show how
and when the two descriptions can be related.

At each point of the curve associate three orthogonal unit vectors
$\{\t,\n,\b,\}$. The vector ${\bf t}$ is tangent to the curve, while
${\bf n}$ and ${\bf b}$ are perpendicular to the tangent, they both
live in the surface of the sphere of fig. (\ref{writhefig2}).  A
theorem of Euler \cite{euler} then shows that the motion of this frame
as a function of $s$, the curvilinear distance, is a pure rotation.
Thus grouping the vectors into a matrix, ${\bf M}$ gives
\begin{equation}
{{d \bf M } \over {ds }} =  
{ \mathbf \Omega}.
{\bf M}
\label{rotate}
\end{equation}
with $\Omega$ anti-symmetric.  Different choices are possible; we
shall now discuss two: firstly the classic Frenet frame,
known from elementary treatments of line geometry in three dimensions,
secondly the choice of parallel transport corresponding to fig. (\ref {writhefig}).

The most common choice for the rotation matrix, is the following
corresponding to the Frenet frame \cite{frenet},
\begin{equation}
\Omega = \left( \begin{array}{ccc} 0 & \kappa & 0 \\
 -\kappa & 0 & \tau \\
 0 & -\tau & 0
\end{array} \right )
\label{frenet} \ .
\end{equation}
This representation is particularly useful because $1/\kappa$ has a
simple geometric interpretation: it is the radius of curvature of the
line at the point $s$. The element $\tau$ is known as the {\it
  torsion}.

As shown above the geometry of three dimensional curves becomes easier
to understand if we work in the tangential space $\s2$.  What is the
geometric picture that corresponds to the choice of the matrix in eq.
(\ref{frenet})?  On the sphere, the curve $\ts$ moves at speed
$\kappa$ in the direction $\n' $ which is parallel to $d\t/ds$.  As a
function of $s$ the vector $\n'$ rotates in the surface at angular
velocity $\omega_{\t}=\tau$. Since we are particularly interested in
the internal geometry of the sphere let us now transform the
parameterization of the curve from unit speed in real space to to a
unit speed curve on the sphere. The element of length transforms as $d
\sigma = \kappa ds$; the angular velocity of $\n'$ expressed as a
function of $\sigma$ (rather than $s$) is $\tau / \kappa$,
\cite{geod}. In fig. (\ref{writhefig2}) $\tau/\kappa$ is zero along the arcs
$\alpha$, $\beta$, $\gamma$ but at the corners, $A$, $B$ and $C$ there
is a rapid rotation of $\n'$ by $90^\circ$, corresponding to a
delta-function singularity in the angular velocity.  The
representation becomes {\it singular}\/ in for curves which are not
sufficiently smooth, or for curves where $\kappa$ passes through zero.

The second representation of the geometry (sometimes known as the
Fermi-Walker frame), \cite{segert} is the following
\begin{equation}
\Omega=
\left (
\begin{array}{ccc}
  0 & \alpha & \beta \\
  -\alpha & 0 & 0 \\
  -\beta & 0 &  0 
\end{array}
\right )
\label{parallelframe} \ .
\end{equation}
It corresponds to the geometry shown in figures (\ref{writhefig}, \ref{writhefig2}).  In this frame
the vectors ${\bf n}$ and ${\bf b}$ are parallel transported: The
original definition \cite{parallel} of parallel transport of a vector
${\bf v}$ (due to Levi-Civita) is that the rate of change of the
vector, projected back on to the tangent plane is zero. For our space
curve this corresponds to the equation
\begin{equation}
{d{\bf  v}\over {d \sigma}} - \t \left({\bf t}. {{d{\bf  v}}\over{d\sigma}}\right)=0.
\end{equation}
We see that both ${\bf n}$ and ${\bf b}$ (as well as an arbitrary,
constant linear combination of $\b$ and $\n$) evolving with the matrix
eq.  (\ref{parallelframe}) obey this equation. In figure
(\ref{writhefig2}) the vector $\n$ is translated in such a way that $d
\n /d\sigma$ is always normal to the surface, in agreement with this
definition.

For sufficiently smooth curves both representations are equivalent.
At each point on the curve perform a rotation about the vector ${\bf
  t}$ in the following manner:
\begin{equation}
\left (
\begin{array}{c} {\bf n'}  \\ {\bf b'}  
\end{array}
\right)
=
\left (
\begin{array}{cc} \cos{\theta} & \sin{\theta} \\ -\sin{\theta} & \cos{\theta} 
\end{array}
\right )
\left (
\begin{array}{c} 
  {\bf n}  \\ {\bf b}  
\end{array}
\right) \ ,
\end{equation}
where ${\bf n'}$ and ${\bf b'}$ are the vectors in the description eq.
(\ref{frenet}) and ${\bf n}$ and ${\bf b}$ are the vector in the
description of eq. (\ref{parallelframe}). The two representations are
be linked by setting
\begin{eqnarray}
\dot \theta &=& \tau \\
\alpha &=& \kappa \cos{\theta}\\
\beta &=& \kappa \sin{\theta} \ .
\end{eqnarray}

Consider an arbitrary trajectory on the sphere comparing the two
frames, starting from a common initial vector $\n (0)$. After the loop
there has been a rotation of
\begin{equation}
\theta_L=\int_0^L \tau \ ds = \int {\tau\over \kappa} \ d\sigma
\end{equation}
between the two description. If the angle between the parallel
transported frames $\n (0)$ and $\n (L)$ is $\Delta_\Omega$ then the
angle between $\n (0)$ and $\n' (L)$ is $(\Delta_\Omega + \theta_L)$.
Clearly this must be equal to a multiple of $2 \pi$ since $\n' (0)$
and $\n' (L)$ are identical (since they are both equal to
\begin{math}
  d \t/d\sigma
\end{math}). Thus we deduce
\begin{equation}
\Delta_\Omega + \int {\tau  \over \kappa}\ d \sigma = 2 n \pi
\label{theorem} \ ,
\end{equation}
which when combined with equation (\ref{gauss}) is the Gauss-Bonnet
theorem for a sphere.

\section{Examples of Writhe}

\subsection{Bent Bar}

The trajectory on the sphere, shown in in fig. (\ref{writhefig2}) has
three rotations of $\pi/2$ corresponding to each bend in the bar thus
$ \int (\tau/ \kappa) \ d\sigma = 3 \pi/2$. $\Delta_\Omega$ is also
equal to $\pi/2$ as is the area enclosed by the curve on the sphere as
required by eq (\ref{gauss}).  Thus $2 \pi \wr + \int \tau \ ds = 2 \pi$.

\subsection{Circular Helix}  
A second example is a long helix, parameterized by the equation
\begin{equation}
{\bf r}(u) = (a \cos{u}, a \sin{u}, u)
\end{equation}
for the case $a$ small. The torsion is given by
\begin{equation}
\tau={1 \over 1+a^2} \,
\end{equation}
with $ds = du \sqrt{1+a^2}$. As a function of $s$ the trajectory of
$\t$ on the sphere is a circle or radius $a$. Consider the various
contributions to eq. (\ref{theorem}) for a line of length
\begin{math}
  L={2\pi N} \sqrt{1+a^2}
\end{math}
, so that the are $N$ complete helical repeats.  The integral of the
torsion is given by
\begin{equation}
\int_0^L {ds \over {1+a^2}}  \approx 2 N \pi - N \pi a^2 +O(a^4)
\label{small} \ .
\end{equation}
The result is at first very surprising. In the limit $a \rightarrow 0$
the helix becomes a straight line, however the torsion and its
integral remains finite: The straight reference state is singular for
the calculation of the torsion of a curve.

This second example shows that the torsion is sensitive to small scale
details in the representation of a curve, in particular to small scale
helical structure (as is found in stiff biomolecules).  At finite
temperature the situation is even more delicate.  Even for a filament
with a linear ground state small scale structure is excited by thermal
fluctuations so that the integral of the torsion is badly behaved.  In
fact the standard relationships between writhe and torsion, 
 lead to large, uninteresting zero order contributions
which mask the interesting higher order term, as found here for a
simple helix.

The enclosed area by the path $\t(s)$, is better behaved than the
torsion so that we can use eq. (\ref{gauss}) and the fact that the
circle of area $\pi a^2$ is traversed $N$ times to find $\Delta_\Omega
= N \pi a^2$. The total rotation, which is physically interesting,
corresponds to the second order term in $a^2$ appearing in eq.
(\ref{small}). We conclude that it is better to calculate the writhe
from Fuller's result on the enclosed area of the curve $\t(s)$, rather
than from the integrated torsion. We now generalize these remarks to
finite temperature.

\subsection{Local Geometry}
Locally a smooth writhing curve can be written in the following form
\begin{equation}
{\bf r}(u) = \left (-{\kappa \tau  \over 6}s^3, {\kappa  \over 2}s^2, s \right) \ ,
\end{equation}
showing that $\kappa \tau$ measures the rate at which the curve
becomes non-planar.  We have chosen local coordinates so that the
tangent points in the $\hat {\bf e}_z$ direction. From this expression
we again see that a straight filament with $\kappa=0$ can be
approached with an arbitrary value of the torsion $\tau$.

On $\s2$ we find that
\begin{math}
  \t \approx (-{\kappa \tau s^2/ 2},\kappa s,1).
\end{math}
When we transform  to unit speed on $\s2$ using $\kappa s=\sigma$ we
find
\begin{equation}
\t \approx \left (-{\tau\over 2 \kappa}\sigma^2, \sigma, 1\right )\ ,
\end{equation}
so that the projection of the path from the sphere onto the tangent
plane is curved, with radius of curvature $\kappa/\tau$. Thus $\tau/\kappa$
has a natural interpretation on the sphere and is indeed the {\sl geodesic curvature}
of the trajectory on $\s2$.

\section{Bending of stiff polymers}
\subsection{Energetics}
The bending energy of a stiff polymer is written in the form
\begin{equation}
E = {K\over 2} \int \left( {d \t (s)\over ds}  \right)^2 \ ds
\label{stiff} \ ,
\end{equation}
where $K$ is the bending modulus of the polymer. Let the energy be
measured in units of the temperature then $K$ is the persistence
length of the polymer.  This energy is completely analogous to that
used in the theory of {\it flexible}\/ polymers \cite{doi}
\begin{equation}
E_{flex} = {\Sigma \over 2} \int \left( {d {\bf r} 
(s)\over ds}  \right)^2 \ ds 
\label{flex} \ ,
\end{equation}
except that it is written in terms of the tangent vector ${\bf t}$
rather than the real space position ${\bf r}$. The energy (\ref{flex})
describes a Gaussian polymer with end to end distance $\langle R^2
\rangle = 3L/\Sigma$. Like a flexible polymer the distribution of
density $P({\bf t},s)$ obeys a diffusion like Fokker-Planck equation
\cite{doi}.
\begin{equation}
{\partial P(\t ,s) \over \partial s}
=
{1\over 2 K}
\nabla^2_{\sigma}P(\t ,s) \ ,
\label{fp}
\end{equation}
where $\nabla^2_{\sigma}$ is the Laplacian operator on the sphere.  As
a function of $s$ the vector ${\bf t}$ diffuses with diffusion
coefficient $1/(2K)$. The typical separation on $\s2$ of two points
separated by a curvilinear distance $s$ in real space is
\begin{equation} R_{\sigma}^2 = 2s/K
\label{fokker},
\end{equation}
when $s \ll K$.
 
From the result eq. (\ref{fp}) we understand why torsion, $\tau$ is
poorly defined at finite temperatures: Consider a discretization of a
semiflexible polymer with sections of length $a$ in real space. On the
sphere the polymer is represented as a random walk with points
separated by a distance comparable to $\sqrt{a/K}$.  The direction of
each step on the sphere is completely {\it uncorrelated}\/ with its
neighbors.  Thus the angle between successive links is of order of one
radian. There are $L/a$ sections so that the torsion locally has a
value $\tau \sim \pm 1/a$ and $\int \tau ds \sim \pm \sqrt{L/a}$ which
diverges with the cutoff. The physically interesting writhe is thus
hidden by the diverging zero order fluctuations. This is similar to
the example of the helix above where the physically interesting writhe
was hidden by the small amplitude helical pitch. We shall now show
that despite some subtleties the angle between the vectors $\n(0)$ and
$\n(L)$ can be calculated from the area enclosed by the random walk on
$\s2$.
 
This result can be linked with the natural class of functions for
which the Frenet frame is defined. Bishop \cite{bishop} showed that
the Frenet frame (and thus the torsion) requires a function space of
${\cal C}^{3}$ curves. The frame defined by parallel transport has
lower smoothness requirements.  Clearly a semiflexible polymer, which
is only $C^{3/2}$, is too rough for the Frenet frame to be useful.

\subsection{Short Polymers}
Consider a polymer with the following boundary conditions: The ends of
the filament ${\bf r}(0)$ and ${\bf r}(L)$ are free to move but the
chain is constrained by external couple so that ${\t}(0)$ and
${\t}(L)$ point in the $\hat {\bf z}$ direction, as in fig. (\ref{writhefig}). The
trajectory is closed on the sphere (but open in real space).  If the
polymer is short, ($L\ll K$) the trajectory never moves too far from
the north pole; locally the geometry of the sphere remains Euclidean.
The writhe of the filament can thus be calculated from the area
enclosed by a closed Gaussian loop in a plane. This problem has been
treated by Levy \cite{levy}.  The enclosed area, fig. (\ref{walk}),  scales
in the same manner as $R^2_{\sigma}$ but is randomly signed (depending
on whether the path is traversed clockwise or anti-clockwise). The
mean squared area has a non-zero average:
\begin{equation}
\langle \wr^2\rangle = {  L^2\over 48 K^2} \ .  
\label{shortwrithe}
\end{equation}
In fact the whole probability distribution is known and is given by
\begin{equation}
{\cal P}_K (\wr) =  {\pi K \over  L } {1 \over \cosh^2(2 \pi\, \wr\,  K/L )}
\quad .
\label{writhedis}
\end{equation}
Although the distribution of the size of the loops is Gaussian the
distribution of the writhe has wings which are simple exponentials,
implying that the free energy of a strongly writhing state varies
linearly with $\wr$,
\begin{equation}
  F=4 \pi K\,\wr /L \, .
\end{equation}
This result implies that there is a critical torque $\Gamma=2K/L$,
which destabilizes fluctuations about the ground state of the polymer.

\begin{figure}[ht]
  \includegraphics[scale=.75] {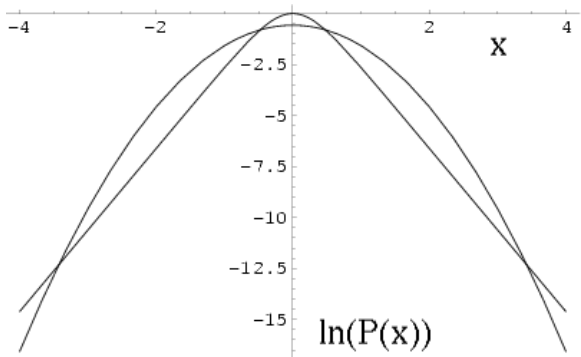}
\caption{{\small 
    Distribution functions eq. (\ref{writhedis}), (\ref{twistdist}) of angle of rotation,$x$ (on a log-linear
    scale) due to writhe and twist.  The parabolic curve is the
    distribution of twist, the curve with the exponential wings is the
    distribution of rotation due to writhe.  Plot is for $T/L =K/L=2$
    } \label{distribution} }
\end{figure}

\begin{figure}[ht]
  \includegraphics[scale=.6] {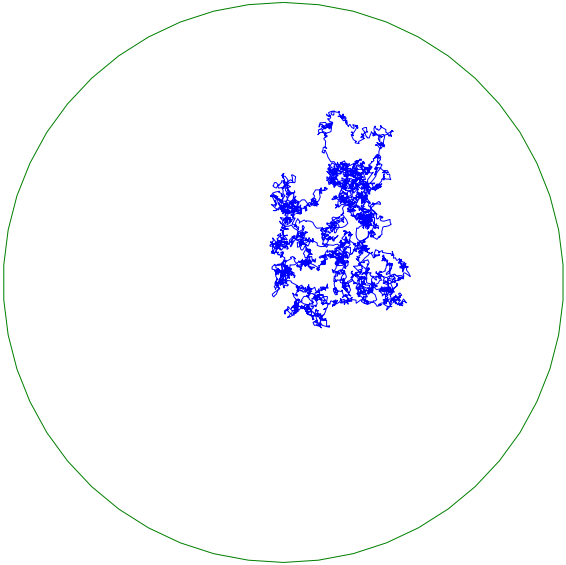}
\caption{{\small Realization of a closed random walk on a sphere. The
    walk has a fractal nature with detail on all length scales. The
    area enclosed by the loop will be dominated by the largest loops
    in the above figure so that the area enclosed scales as the radius
    of gyration squared.} \label{walk}}
\end{figure}

For an elastic filaments with torsional elastic constant $T$ the total
angle of rotation between two ends of a polymer is given by the sum of
the writhing and twisting contributions.  The twisting contribution,
$\theta$ is Gaussian distributed:
\begin{equation}
{\cal P}_T(\theta) =\sqrt{T\over{2 \pi L}} e^{-\theta^2 T/2L}\ .
\label{twistdist}
\end{equation}
We can find the distribution of the total rotation angle $\varphi$ from a convolution
of the writhe and twist distributions
\begin{equation}
{\cal P} (\varphi) = \int {\cal P}_K( \wr )\, {\cal P}_T({\theta})\,  \delta( 2 \pi Wr + \theta - \varphi) \ d\wr \ d\theta \ .
\end{equation}
The twisting and writhing distributions are plotted in fig.
(\ref{distribution}).

\subsection{Long Polymers}
The writhe properties of long semiflexible polymers have been
extensively studied in the context of DNA super-coiling
\cite{russians,frank,marko}. It is known from numerical simulations that for
long semiflexible polymers $\langle \wr^2 \rangle \sim L$. Recently,
detailed analytic calculations \cite{mezard} have shown that this is
only true in models with a microscopic cut off. In the absence of a
cut off the writhe fluctuations are logarithmically divergent.  How do
we understand these results from our picture of a stiff polymer as a
random walk on the sphere?

As shown by Fuller \cite{fuller} the writhe of an arbitrary
configuration of a polymer is found from the expression
\begin{equation}
\wr = {1\over 2 \pi} \int {(\t_0 \times \t)\over 1+ \t .\t_0}. \ d\t
\label{straight}
\end{equation}
where $\t_0$ is an arbitrary constant vector. This expression is valid
when the path $\t(s)$ can be deformed from the vector $\t_0$ without
the denominator vanishing. Let $\t_0$ point in the direction of the
north pole then the singularity is at the south pole.  To understand
the nature of the singularity expand about the south pole so that
\begin{equation}
\t \approx (u,v,-1+u^2/2 +v^2/2)
\end{equation}
with both $u$ and $v$ small.  The contribution to the writhe coming
from a section of polymer near the south pole has the following form
\begin{equation}
\Delta \wr \approx {1\over  \pi} \int  {u\, dv -v  \, du\over {u^2 +v^2}} 
 = {1\over  \pi}\int d \phi \ ,
\end{equation}
where $\phi$ is the azimuthal angle of the path subtended at the south
pole. The contribution of the polymer to the writhe near the pole is
twice the winding number about the pole. This winding number is
clearly discontinuous as a function of the position of a path: A small
circle centered at the pole has winding number unity. If it is
displaced more than its radius the winding number jumps to zero.

We conclude that small fluctuations in the direction of the polymer
can give important ($O(1)$) fluctuations in the total writhe if they
occur near the south pole. The distribution of winding of a random
walk on a sphere has been given analytically \cite{spitzer,orsay}. It
is known that the distribution is wide and has a Cauchy tail.  This is
due to the fact that in a close approach to the pole to within a
distance $\epsilon$ there are typically $\sqrt{\log{\epsilon}}$
rotations about the pole.
This is the origin of the Cauchy tail in the writhe fluctuations which decays
even more slowly than the curve plotted in fig. (\ref{distribution})
\cite{mezard}.

Is this singularity at the south pole important for short polymers?
When the polymer is very short transit in the region of the south pole
is strongly suppressed since a bend of $180^\circ$ costs an energy
$\pi^2 K/2L$ leading to exponentially small corrections to the law
(\ref{shortwrithe}).  However as soon as $L\sim K$ the dense nature of
random walk in two dimensions means that with high probability there
is a point on the polymer which is very close to the south pole,
giving a strong Cauchy tail to the writhe distribution.

\section{Writhing dynamics}

When studying the dynamics of writhing of stiff polymer the
representation of the writhe given by eq. (\ref{straight}) is
only useful for filaments which are much shorter than the persistence
length. Motion of a long polymer is almost surely going to cause
passage of the path over the south pole leading to a discontinuous
contribution to the writhe. It is better to use the generalization
\cite{fuller} of eq. (\ref{straight})
\begin{equation}
\wr(\t_2) - \wr(\t_1) = 
 {1\over 2 \pi} \int {{ \t_1 \times  \t_2}\over {1+\t_1.\t_2}}
\; .(\dot \t_1+\dot \t_2)\ ds  \ .
\end{equation}
Let the trajectory $\t_1(s)$ to be the polymer at
$t=0$ and $\t_2(s)$ is the trajectory at some later time.  This
expression remains valid until a time such that a point on $\t (s)$
rotates $180^\circ$ leading to a singularity in the integral.
 
The motion of a semiflexible filament is given by the Langevin
equation \cite{zimm}
\begin{equation}
4 \pi \eta {\partial  {\bf r_{\perp}} \over \partial t} 
=
-K  {\partial^4 {\bf r_{\perp}}\over \partial s^4} + {\bf f_{\perp}} (t,s) 
\ .
\end{equation}
This equation has a characteristic dispersion relation $\omega \sim
q^4$ so that after a time $t$, wavelengths out to 
\begin{math}
l_1 = (K
t/\eta)^{1/4}
\end{math} are equilibrated. Perturbations separated by more than the
distance $l_1$ have not had time to interact so that there are
$N_{1}= L/l_1$ dynamically independent sections remaining on the
polymer.

What does this mean for the structure of the path $\t (s)$ on $\s2$?
From eq. (\ref{fokker}) the length $l_1$ in real space translates
to a distance $R_{1} \sim \sqrt{l_1/K}$ on $\s2$. Structure on
the scale of $R_{1}$ has had time to move.  We can thus
imagine the structure at time $t$  as being a series of
$N_{1}$ beads of size $\sqrt{l_1/K}$ strung out along the original
path of the polymer on $\s2$. The typical fluctuation in area
 from each bead is given by
\begin{math}
  (\Delta A)=\pm R_{1}^2
\end{math}.
To find the total fluctuation in area in time $t$  add these
$N_{1}$ independent fluctuations to find
\begin{eqnarray}
\langle (\wr(t) -\wr(0))^2 \rangle &\sim& N_{1} (\Delta A)^2 \sim L l_1 /K^2\\
& \sim & L t^{1/4} \ . 
\end{eqnarray}
This result has been found recently in detailed numerical simulations
\cite{me}.  The derivation in terms of random walks shows that the
result is valid for polymers much longer than the persistence length:
Indeed the result is valid for all times such that $l_1(t) \ll K$.
Beyond this time there are two sources of breakdown in the theory:
Firstly the dynamics are no-longer described by a simple bending
theory and secondly we also expect that rotations by $180^\circ$
become important leading to a breakdown of our expression for the
writhe.  

Since the writhe fluctuations are the sum of a large number of
independent contributions dynamic writhe fluctuations are distributed
according to a Gaussian law. However there is always a exponentially
small probability that the singularity in the integral occurs very
rapidly so that the expressions found here is only be
asymptotic, as was the case for the {\it static}\/ distribution of
writhe for short filaments eq. (\ref{writhedis}) found above.

\section{Writhing of bent polymers}
The discussion until now has been restricted to polymers
which are straight in their ground states. It is known, however, that certain
special sequences of DNA lead to ground states which are strongly
curved. This leads to enhanced writhing.
 Consider a polymer  which bends an angle $\Psi$
over a distance $L_b$ in its ground states.
On the sphere the ground state is a geodesic of length
$\Psi$.  We have seen above that a straight polymer is
described by a diffusion like Fokker-Planck equation on the sphere.
Addition of drift due to the spontaneous curvature leads to the
equation
\begin{equation}
{\partial P(\t ) \over \partial s}
=  { \Psi \over L_b} {\bf v}. \nabla P +
{1\over 2K}
\nabla^2_{\sigma}P(\t ) \ ,
\label{fp2}
\end{equation}
with {\bf v } a unit vector describing the direction of the bend.  We
take as our reference configuration $\t_1$ the ground state and the
state $\t_2$ the thermalized polymer.  As shown above the typical
fluctuations of the filament on $\s2$ scale as 
\begin{math}
 R_{b} \sim  \sqrt{L_b/ K}
\end{math}
so that the  area enclosed  by the pair of paths $\t_1$ and $t_2$ scales as
\begin{math}
\pm \Psi
R_{b} 
\end{math}. Thus we find that 
\begin{equation}
\langle \wr^2\rangle \sim {L_b  \Psi^2 \over K} \ .
\end{equation}
This implies that a polymer, such as DNA with a sharp hairpin is
expected to display enhanced torsional fluctuations at its ends
due to this amplification of the writhe fluctuations by the
spontaneous curvature. 

Similar modifications occur in the writhe dynamics: As
before there are $N_{1}=L_b/l_1$ dynamically independent sections
on a polymer, each section moves a distance $\sqrt {l_1/K}$. It is stretched
a distance $ \Psi l_1/L_b$ by the drift. Thus writhe fluctuations scale
as
\begin{eqnarray}
\langle (\wr(t) -\wr(0))^2 &\sim& N_{1} (( l_1  \Psi/L_b) 
 \sqrt {l_1/K})^2 \\
&\sim& t^{1/2} \Psi^2/L_b \ ,
\end{eqnarray}
showing that this effect is most important in short, sharp bends.

\section{Writhing of closed loops}
Recently simulations have been performed \cite{tamar} to study the
writhing dynamics of short closed DNA loops.  Closure of the path in
real space corresponds to adding a global constraint on the tangent
\begin{equation}
\int \t(\sigma) \ {d\sigma \over \kappa} = 0
\label{constraint}
\end{equation}
for all times. Here we consider the effect of this constraint on the statics
and dynamics of writhe fluctuations for a short polymer such that $L
\ll K$. 

The zero temperature configuration of the polymer on $\s2$ is a great
circle which we take to be the equator. From eq.  (\ref{constraint})
the average vertical position on the sphere can not change: Any motion
of the polymer in the north-south direction must be compensated by a
corresponding south-north motion elsewhere in the polymer so that the
average vertical position of the polymer does not move. However on the
sphere this is a {\it mass weighted}\/ average with the ``mass
density'' $1/\kappa$ (Note: The enclosed area is calculated with a unit
weighted measure).  By coupling fluctuations in density along the
equator with transverse fluctuations in the position of the equator we
can find a non zero contribution to the writhing.

The lowest order fluctuation in longitudinal density which does not
violate the constraint of eq. (\ref{constraint}) is proportional
to $\cos{2 \phi}$ where $\phi$ is the azimuthal angle.  From eq.
(\ref{fp2}) its amplitude scales as $\sqrt{L/K}$.  Coupling this
density fluctuation to the transverse motions, also of magnitude
$R_{\sigma}=\sqrt{L/K}$ leads to a typical writhe fluctuation of $\pm
L/K$ for a closed loop. Thus we find that
\begin{equation}
\langle \wr^2 \rangle \sim L^2/K^2.
\end{equation}

How does the closure constraint modify the writhing dynamics?  The
geometry is very similar to that discussed above for the bent open
filament. There are $N_{1} =N/l_1$ dynamic sections of length $2 \pi
l_1/L$ moving a distance $\sqrt{l_1/K}$. Again the first order
contribution must vanish, since the ``mass weighted'' position of the
equator can not move. A non-zero term can be
generated by coupling to the longitudinal density fluctuations of
amplitude $\sqrt{L/K}$:
\begin{equation}
\langle (\wr(t)-\wr(0))^2 \rangle \sim l_1^2 /K^2 \sim t^{1/2}\ .
\end{equation}
The absolute magnitude to the fluctuations is thus decreased compared
with the open polymer but the dynamic exponent for the fluctuations
remains $1/2$. Recently \cite{tamar} molecular dynamics simulations
were performed on a detailed microscopic model of DNA loops. The
numerical results seem to be consistent with a square root of time
evolution of the writhe-writhe correlation function.

\section{Open Filaments}
\subsection{Relation with geometric phases}

The geometry of fig. (\ref{writhefig}) is identical to that used to
discuss the propagation of polarized light along bent fibre optics
\cite{berry,haldane,optics}: Consider a circular cross section optical
fiber bent into the shape shown in fig. (\ref{writhefig}).  If plane
polarized light is sent down the fiber with the plane of polarization
defined by the vector ${\bf n}$ then the polarization will be
transported in exactly the same manner as the normal vector to the bar
\cite{segert}. After the triple bend of fig. (\ref{writhefig}) the
plane of rotation of the light has been rotated by $90^\circ$. The
rotation of the plane of rotation in fibre optics has been extensively
studied as a simple example of the Berry or geometric phase. Berry's
results on the existence of non-trivial phase factors is
mathematically related to Fuller's theorem for writhe.  While the
geometry of the two situations seems very similar the geometric phase
corresponding to this experimental situation is somewhat disguised so
we shall now show how the rotation of the plane of polarization of
light is equivalent to a phase shift for a photon.

There are two equivalent mathematical descriptions available for
polarized light: Firstly, planar polarized waves, secondly circularly
polarized photons. Using as a basis the planar representation with
eigenstates $\phi_h=(1,0)$ and $\phi_v=(1,0)$, the helical eigenstates
are
\begin{math}
  \phi_r= {1\over \sqrt{2}}(1,i)
\end{math}
and
\begin{math}
  \phi_l= {1\over \sqrt{2}} (1,-i)
\end{math}. Both the pair $\{ \phi_v, \phi_h \}$ and the pair $\{
\phi_l, \phi_r \}$ form a complete orthonormal basis for describing
arbitrary pure states of the light.  Consider and experimental set up
that rotates the vector $\phi_h$ by an angle $\psi$. The new state is
given by $(\cos{\psi},\sin{\psi})$.  This state can also be expressed
as a linear combination of the circularly polarized states.
\begin{equation}
2\left (
\begin{array}{c}
  \cos{\psi} \\
  \sin{\psi}
\end{array} \right)
=
\exp(-i \psi )
\left(
\begin{array}{c}
  1 \\
  i
\end{array}
\right)
+
\exp(i \psi )
\left(
\begin{array}{c}
  1 \\
  -i
\end{array}
\right) \ .
\end{equation}
The rotation of the plane of polarization is due to the modification
of the relative phases of the two helical states.  One state is
advanced by $\psi$ the second is retarded $\psi$.  When a beam of
circularly polarized light is sent down a fiber there is a phase shift
which is equal to the writhe of the fibre times the helicity of the
photon.  This is a Berry phase.

A treatment of the geometry of writhe via a phase leads to an
immediate generalization of Fuller's theorem for trajectories which
are not closed on the sphere \cite{panch,haldane}: The rotation angle
between $\n (0)$ and $\n (L)$ is for the moment undefined if $\t(0)$
and $\t(L)$ are not parallel. This is because we have only 
  defined parallel {\it locally} on $\s2$ via parallel transport. We shall
now see how find a {\it global} definition of parallel by noting that
the phase phase difference between two helical states is unambiguous
even if the trajectory does not close on $\s2$: The phase shift in a
fiber with arbitrary boundary conditions corresponds to the following
recipe for calculating the writhe of the filament \cite{panch}: Follow
the path $\t(0)$ to $\t(L)$ then close the path from $\t(L)$ to
$\t(0)$ by a great circle (or geodesic). The phase shift is given by
the area enclosed by the path $\t(0) \rightarrow \t(L)$ augmented by
the great circle $\t(L) \rightarrow \t(0)$.

\begin{figure}[ht]
  \includegraphics[scale=.30] {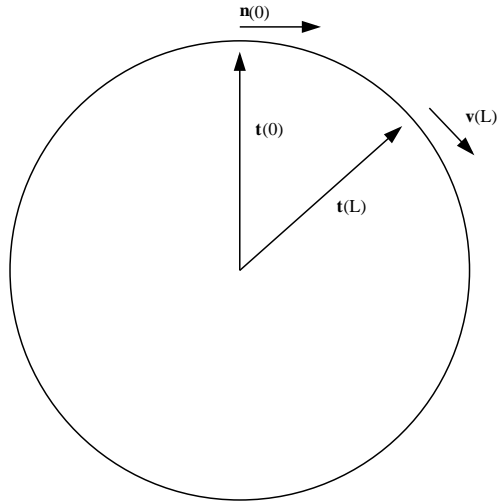}
\caption{
  \small { Geometry of an open trajectory on $\s2$. The vectors
    $\t(0)$, $\t(L)$ and $\n(0)$ can be chosen to be co-planar. The
    vector ${\bf v}(L)$ is a reference vector at the point $s=L$ which
    allows one to compare rotations with $s=0$. The ``most parallel''
    choice possible for ${\bf v}(L)$ is shown in the figure, and is
    coplanar with the other vectors in this figure.  }  }\label{close}
\end{figure}

This recipe, fig. (\ref{close}) has a interpretation on the sphere $\s2$, for the writhing
of a stiff polymer. Consider a path $\t(0)$ to $\t(L)$.  The choice of
$\n(0)$ is arbitrary for the calculation of the rotation, for
convenience let us choose $\n(0)$ to that it points towards $\t(L)$.
The three vectors $\t(0)$, $\t(L)$ and $\n(0)$ define a plane.  Choose
a reference vector, ${\bf v}(L)$, on the sphere at $\t(L)$ so that it
is the ``most parallel'' possible to $\n(0)$.  We choose ${\bf v}(L)$ so
that it is coplanar with $\t(0)$ $\t(L)$ and $\n(0)$. This {\it
  definition} of {\it parallel} for two points on the sphere separated
by a finite distance allows a definition of writhe for arbitrary free
boundary conditions, leading to the geodesic recipe for calculating the
writhe.

\subsection{Perturbative treatment of open filaments}

For short filaments there is an alternative derivation of the result
that the path on the sphere must be closed by a geodesic. We shall use
standard techniques from time dependent quantum mechanics, valid in
the limit $L/K \ll 1$.  We proceed by studying eq. (\ref{rotate}) for
the case of a general matrix, $\Omega(s)$, describing both bend and
twist, with ${\bf \Omega}_{i,j}=\epsilon _{ijk}\omega^k$.  If the
average direction of the polymer is aligned in the $\hat {\bf e}_z$
direction the angular velocities $\omega^x$ and $\omega^y$ correspond
to bending the filament, $\omega^z$ is the twist.  We iteratively
integrate this equation.  The lowest order writhing contribution to
rotations about the z-axis comes in second order.
\begin{equation}
  \Omega(\wr) = {1\over 2}\int_0^L  ds ds' \  \theta(s-s')\ (\omega_s^x \omega_{s'}^y - \omega_{s'}^x \omega_s^y).
\end{equation}
This is just the time ordered commutator of the rotation operators
familiar from quantum perturbation theory.  Integrating by parts to
transform the $\theta$-function into a $\delta$-function gives
\begin{equation}
\Omega(\wr) = {1\over 2} \hat {\bf e}_z.\left ( \int_0^L ( \dot {\bf t} \land {\bf t} )ds + {\bf t}(0) \land {\bf t}(L) \right)
\end{equation}
For a closed curve on $\s2$ the first term is the area enclosed by the
curve ${\bf t}(s)$ on a patch of a sphere in agreement with Fuller's
theorem: It is the first term in the expansion of eq.
(\ref{straight}).  The second boundary term, due to the integration,
present when the curve is open corresponds to closing the path by a
geodesic from ${\bf t}(L)$ to ${\bf t}(0)$.

\subsection{Dynamics of open filaments}
This definition of writhe has interesting consequences on the
measurement of writhe fluctuations on freely fluctuating filaments.
When the ends of the filament were constrained we showed that the
writhing was due to a series of $N_{1}$ small fluctuations distributed
along the chains.  If $\t(L)$ can also fluctuate there is a new
contribution due to the modification in the closing of the path. In a
time $t$ $\t(L)$ moves $\sqrt{l_1/K}$ sweeping out an area
$\sqrt{l_1/K} \times \sqrt{L/K}$ thus there is an extra contribution
to the writhing dynamics which scales
\begin{math}
  \langle \wr ^2 \rangle \sim l_1 L/K^2
\end{math}
as before but coming from a single large end event rather than the sum
of independent contributions.

\section{Conclusion}

We have shown that the writhing properties of a stiff polymer can be
calculated from the well known properties of Gaussian random walks on
a sphere. The predictions that $\langle \wr^2 \rangle \sim L^2$ and $
\langle (\wr(t) -\wr(0))^2 \rangle \sim t^{1/4} L$ for short filaments
should be accessible to micromanipulation techniques on actin
filaments, or perhaps DNA, either by studying the rotational dynamics
of the end of a marked polymer or by direct observation of the shape
fluctuations occurring in three dimensions via confocal microscopy. As
pointed out in recent work \cite{nelson} disorder can sometimes have
strong effects on the dynamics of twist and torsional modes. It would
be interesting to have a better understanding of the effect of
disorder on the correlation functions discussed here.

All real biopolymers have a helical structure with anisotropic bending
elasticity. It is quite easy to treat this problem in the present
description of diffusion on a sphere by introducing anisotropic
diffusion coefficients on the sphere.

{\small \vskip 0.5cm I wish to thank P. Olmsted for discussions on the
  relationship between geometric phases and Fuller's theorem, N.
  Rivier for discussions on Fermi-Walker (or parallel) transport, D.
  Beard for sending me his raw simulation on writhe fluctuations of
  DNA loops, A. Lesne for references on the distribution of winding
  angles.  }

 \end{document}